\documentclass[%
 reprint,
 amsmath,amssymb,
 aps,superscriptaddress,
]{revtex4-1}

\usepackage{graphicx}
\usepackage{dcolumn}
\usepackage{bm}

\usepackage{comment}
\usepackage{xargs}  
\usepackage{multirow}
\usepackage{xcolor}
\usepackage{braket}
\usepackage[caption=false]{subfig}
\usepackage[normalem]{ulem}
\usepackage{wrapfig}

\usepackage{bibunits}

\let\clearpage\relax
\newcommand{\topic}[1]{{}}

\newcommand{\Ein}{\ensuremath{\eta_{\text{i}}}}
\newcommand{\Eout}{\ensuremath{\eta_{\text{o}}}}
\newcommand{\Eeff}{\ensuremath{\eta_{\text{e}}}}

\newcommand{\blue}{black}
\newcommand{\sqzrotationforSQL}{\ensuremath{\textcolor{\blue}{35^\circ}}}  
\newcommand{\sqzrotationforSRCL}{\ensuremath{\textcolor{\blue}{8^\circ}}} 

\newcommand{\nlgforSQL}{\textcolor{\blue}{4.4$\pm$0.1}} 
\newcommand{\measuredSQZforSQL}{\textcolor{\blue}{3.3 dB}} 
\newcommand{\measuredASQZforSQL}{\textcolor{\blue}{7.7 dB}} 

\newcommand{\circulatingarmpower}{\ensuremath{{\color{\blue}200{\pm}10}}} 
\newcommand{\armpoweruncertainty}{\textcolor{\blue}{5\%}} 
\newcommand{\quantumnoiseatdip}{\textcolor{\blue}{28\%}} 
\newcommand{\dipfrequency}{40 Hz} 
\newcommand{\coatingandthermoopticnoise}{\textcolor{\blue}{24\%}} 

\newcommand{\subsqlpercent}{\textcolor{\blue}{70\%}}
\newcommand{\subsqldB}{\textcolor{\blue}{3 dB}}
\newcommand{\subsqlratio}{\textcolor{\blue}{1.4}}
\newcommand{\sqlrelativetotal}{\textcolor{\blue}{7.2}}

\newcommand{\numberofmeasurementsincontour}{\textcolor{\blue}{12}} 
\newcommand{\statisticalerrorinqnpowerspec}{\textcolor{\blue}{8\%}}
\newcommand{\binwidthforpowerspectrum}{\textcolor{\blue}{0.5 Hz}} 

\newcommand{\calibrationuncertainty}{\textcolor{\blue}{$\pm$3\%}}

\newcommand{\figsqzrotno}{\textcolor{\blue}{3 }}

\begin{document}


\title{Quantum correlations between the light and kilogram-mass mirrors of LIGO}



\overfullrule 0pt 
\parskip0pt
\hyphenpenalty9999


\author{Haocun~Yu}
\affiliation{LIGO, Massachusetts Institute of Technology, Cambridge, MA 02139, USA}
\author{L.~McCuller}
\affiliation{LIGO, Massachusetts Institute of Technology, Cambridge, MA 02139, USA}
\author{M.~Tse}
\affiliation{LIGO, Massachusetts Institute of Technology, Cambridge, MA 02139, USA}
\author{L.~Barsotti}
\affiliation{LIGO, Massachusetts Institute of Technology, Cambridge, MA 02139, USA}
\author{N.~Mavalvala}
\affiliation{LIGO, Massachusetts Institute of Technology, Cambridge, MA 02139, USA}

\author{J.~Betzwieser}
\affiliation{LIGO Livingston Observatory, Livingston, LA 70754, USA}
\author{C.~D.~Blair}
\affiliation{LIGO Livingston Observatory, Livingston, LA 70754, USA}
\author{S.~E.~Dwyer}
\affiliation{LIGO Hanford Observatory, Richland, WA 99352, USA}
\author{A.~Effler}
\affiliation{LIGO Livingston Observatory, Livingston, LA 70754, USA}
\author{M.~Evans}
\affiliation{LIGO, Massachusetts Institute of Technology, Cambridge, MA 02139, USA}
% ...


\begin{abstract}
\end{abstract}

\maketitle

\def\url#1.{}
\newcommand{\urlprefix}[0]{}

\newcommand{\Kappa}{\ensuremath{\mathcal{K}}}

{\bf
Measurement of minuscule forces and displacements with ever greater precision encounters a limit imposed by a pillar of quantum mechanics: the Heisenberg uncertainty principle. A limit to the precision with which the position of an object can be measured continuously is known as the standard quantum limit (SQL)~\cite{CavesPRD1981,SQL1, SQL2, KLMTV}. When light is used as the probe, the SQL arises from the balance between the uncertainties of photon radiation pressure imposed on the object and of the photon number in the photoelectric detection. The only possibility surpassing the SQL is via correlations within the position/momentum uncertainty of the object and the photon number/phase uncertainty of the light it reflects \cite{Unruh1982}. Here, we experimentally prove the theoretical prediction that this type of quantum correlation is naturally produced in the Laser Interferometer Gravitational-wave Observatory (LIGO). Our measurements show that the quantum mechanical uncertainties in the phases of the 200 kW laser beams and in the positions of the 40 kg mirrors of the Advanced LIGO detectors yield a joint quantum uncertainty a factor of \subsqlratio\ (\subsqldB) below the SQL. We anticipate that quantum correlations will not only improve gravitational wave (GW) observatories but all types of measurements in future.
}

\topic{ The HUP and limits to precision:}
The Heisenberg uncertainty principle dictates that once an object is localized with sufficient precision, the momentum of that object must become  accordingly uncertain. In a one-off measurement, this does not pose a problem. But in the case where the position of an object must be measured continuously, as in gravitational wave (GW) detectors, the momentum uncertainty from the act of measuring position evolves into position uncertainty for future position measurements -- a process known as quantum backaction. In striking a balance between the precision of position measurements and the imprecision caused by quantum backaction, an apparent maximum precision for a continuous position measurement is reached. This is the SQL, and for an interferometric measurement, as long as the shot noise and QRPN are uncorrelated, the SQL is indeed the limit.

\topic{ Theoretical paths to beating SQL:}
The SQL was first introduced by Braginsky et al.~\cite{SQL1,SQL2} as a fundamental limit to the sensitivity of gravitational wave detectors. It should be possible to reach the SQL with objects that are {\it macroscopic} or even human-scale, because it is the quantization of the probe light that enforces the SQL (see, e.g., footnote 1 of \cite{KLMTV}). In principle, the SQL can be surpassed when the shot noise and QRPN are correlated. Such correlations already exist in the interferometer, because incoming quantum fluctuations entering from its output port drive both the shot noise and the QRPN, giving rise to ponderomotive squeezing. An injected squeezed state, when combined appropriately with ponderomotive squeezing, enables surpassing the SQL (see Sec. IVB of \cite{KLMTV}). Alternative methods for surpassing the SQL are presented in \cite{KLMTV}, and extended to include optical spring effects in \cite{BnC1}.

\topic{ Beating the SQL experimentally:}

Here, we inject a laser mode in a squeezed vacuum
state in a laser interferometric GW detector with 40 kg mirrors, and use the optomechanically-induced correlations of
ponderomotive squeezing to surpass the free-mass SQL. This measurement marks two significant milestones of quantum measurement. First, we directly observe QRPN contributing to the motion of kg-mass objects, providing evidence that quantum backaction imposed by the Heisenberg uncertainty principle persists even at human scales.    
Second, we surpass the SQL, proving the existence of quantum correlations involving the position uncertainty of the 40 kg mirrors. 
This measurement is an important step
toward further improvements in GW sensitivity through quantum engineering techniques~\cite{KweePRD2014,KLMTV,BnC1,Danilishin2008,PurduePhysRevD2002,polzikBAE2017}.

\topic{ Why the experiment is hard:}
A significant barrier to revealing quantum correlations between light and macroscopic objects is the ubiquitous presence of thermal fluctuations that drive their motion. Previous demonstrations of QRPN have involved cryogenically pre-cooled, pico- to micro-gram scale mechanics~\cite{qrpBoulder2013,schwabBAE2014,wilsonNature2015,teufelPRL2016,polzikBAE2017}, with two exceptions~\cite{cripeNature2019,sudhirPRX2017}. Similarly, previous sub-SQL measurements of displacements have also been performed on cryogenically pre-cooled mechanical oscillators at the pico- \cite{teufelNatureNano2009} to nano-gram \cite{MasonNaturePhysics2019} mass scale. The present measurements are performed on the room-temperature, 40 kg mirrors of Advanced LIGO using 200 kW of laser light, and are enabled by injection of squeezed states and subtraction of classical noise to reveal quantum noise below the SQL. 

\label{sec:Experiment}

\topic{ Experiment overview:} We performed this experiment using the Advanced LIGO detector
in Livingston, Louisiana. For the third astrophysics observing run, squeezed vacuum is injected into the interferometer with squeezing level and squeezing quadrature angle
tuned to maximize the GW sensitivity~\cite{O3Squeezing}. In this experiment, the
interferometer is maintained in the observing configuration ~\cite{instrO3paper}, except data is taken with an increased squeezing level and over a range of squeezing angles, in order to fully
characterize the quantum noise.

\begin{figure}[h]
\centering
\includegraphics[width=0.4\textwidth]{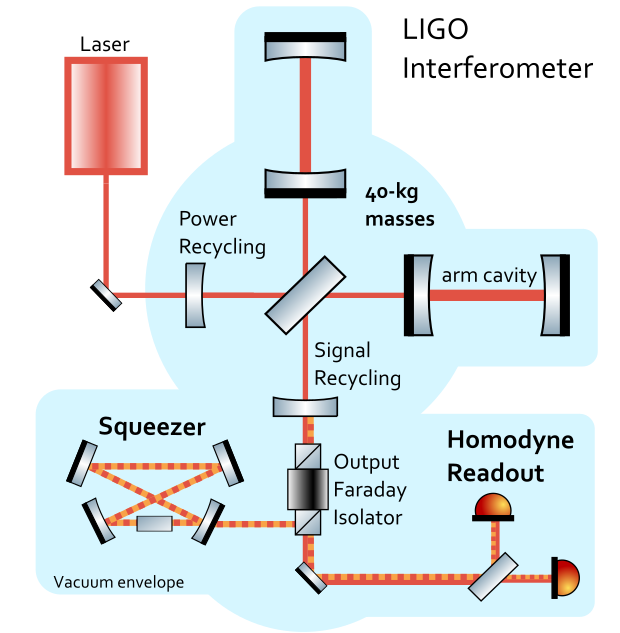}
\caption{Simplified schematic of the experimental setup. Squeezed vacuum (dotted red) is injected through the output Faraday isolator, and co-propagates with the 1064 nm light (solid red) of the main interferometer. A frequency-shifted control field (orange) is used to sense and tune the squeeze angle.}
\label{fig:instrument_layout}
\end{figure}

\topic{ IFO and squeezer description:}
The Advanced LIGO detector is a Michelson interferometer with two 4-km
Fabry-Perot arms, as well as power- and signal- recycling cavities at the input and output ports of the beamsplitter, respectively (see Fig.~\ref{fig:instrument_layout}). The arm-cavity optics are 40 kg fused silica mirrors, suspended as pendulums inside an ultrahigh vacuum envelope \cite{O1instrPRL2016}. During the measurement, \circulatingarmpower\ kW of 1064 nm laser power circulates in each arm cavity. After passing through an output mode cleaner, the differential arm displacement signal ($\Delta x$) is detected as modulations of a small static field at the output due to a deliberate mismatch in the interferometer arm lengths \cite{O1instrPRL2016}. The displacement signal $\Delta x$ is part of a closed servo loop, which is monitored by a continuous calibration procedure that also extracts the instrument sensing function by driving $\Delta x$ motion and measuring the optical response. Details of the squeezed light source and its operation, including the control method for adjusting squeezing angle, are found in~\cite{O3Squeezing}. 
For this measurement, injected squeezing results in \measuredSQZforSQL\ of squeezing and \measuredASQZforSQL\ of antisqueezing measured at the GW readout.

\begin{figure*}[bth]

\centering
\includegraphics[width=0.86\textwidth]{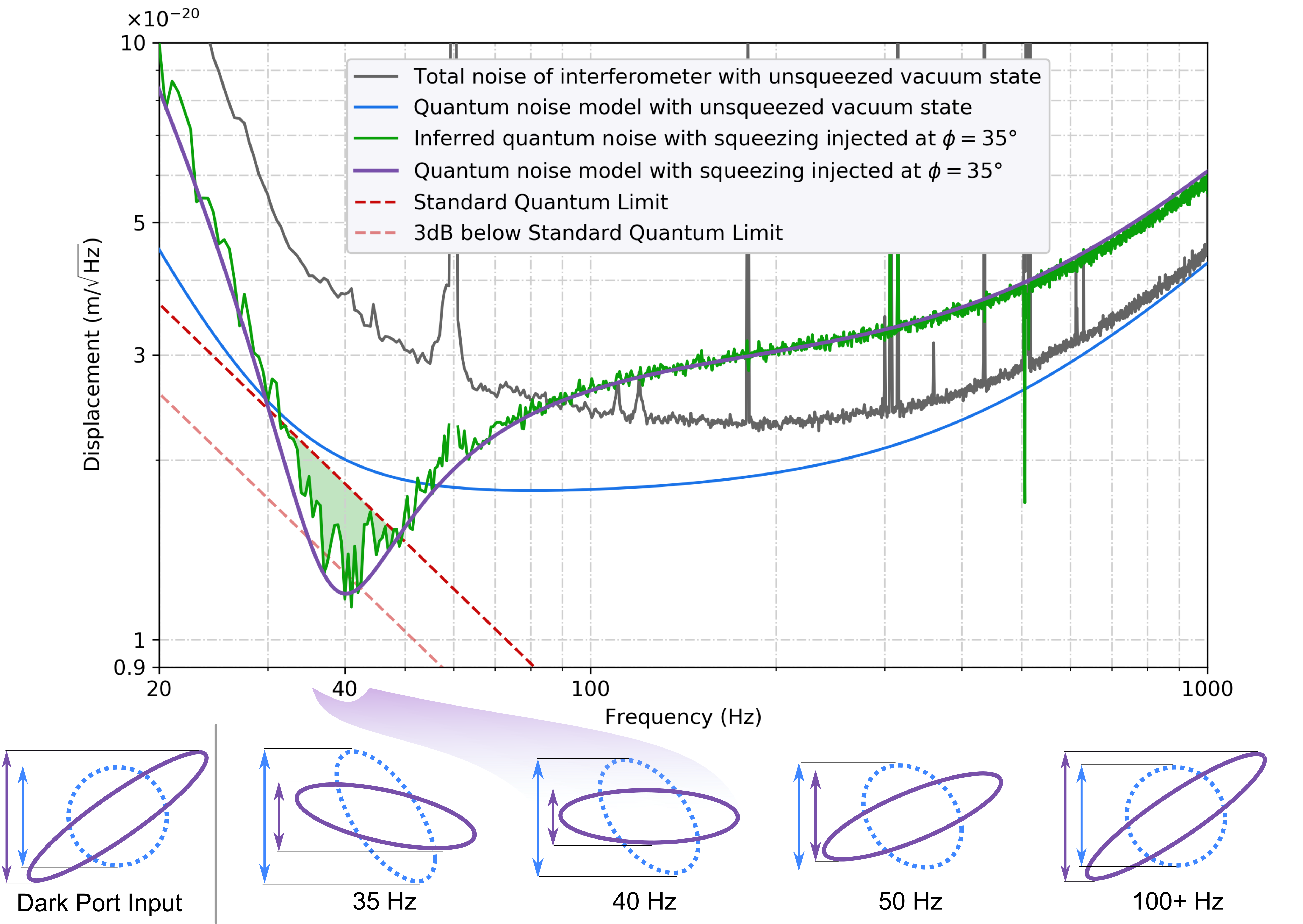} 
\caption{{\bf Top: Differential displacement ($\Delta x$) noise spectral density
of the interferometer.} The grey trace shows the measured total noise level of
the interferometer with unsqueezed vacuum state (i.e. the reference). The blue
trace is the model of quantum noise during the reference measurement. The green
trace shows the inferred quantum noise of the interferometer with injected
squeezing at \sqzrotationforSQL, and its corresponding model is the purple
trace. The notch feature, or ``dip,'' results from the
ponderomotive squeezing affecting the injected optical
squeezed states. It reaches -\subsqldB\ of the free-mass SQL (red dashed
trace, given by Eqn.~\ref{eq:freemass_SQL}) at \dipfrequency. Bottom:
Phase-space representation of the modeled quantum states entering through the
dark port of the interferometer (left) and the output states (right), which are indexed to indicate their frequency dependence. Drawn are the
unsqueezed vacuum (dotted blue) and squeezing at $\phi{=}\sqzrotationforSQL$
(solid purple). In the unsqueezed vacuum case, ponderomotive squeezing distorts the ellipse for frequencies below 100 Hz, increasing QRPN in the readout quadrature (blue arrows). In the injected squeezing case, the same physical process creates a state with reduced noise at \dipfrequency ~(purple arrows).
}
\label{fig:sub-SQLQRPN}
\end{figure*}

\topic{ Model:}
An analytic model of the displacement sensitivity in an idealized LIGO interferometer
illustrates how the combination of ponderomotive squeezing and injected
squeezing allows us to surpass the SQL. A model which builds on methods developed
in~\cite{KLMTV,BnC1}, with extensions to account for losses and off-resonance
cavities, is provided in the Methods section. Here, the idealized model is used for clarity. The Heisenberg uncertainty principle applied to interferometric measurement of differential displacement, $\Delta x$, sets a limit to the one-sided spectral density of:
%
    \begin{align}
       \Delta x^2(\Omega)  &= S(\Omega, \phi)(1 + \mathcal{K}^2(\Omega))\frac{\hbar c }{8k |G(\Omega)|^2 P_{\text{arm}} }
        \label{eq:QL_full}
    \end{align}
    with
    \begin{align}
      \mathcal{K}(\Omega) &= \frac{32k |G(\Omega)|^2P_{\text{arm}}}{m \Omega^2 c}
      &
    G(\Omega) &\equiv \sqrt{\frac{\gamma c}{2L}}\frac{1}{\gamma_{} + i\Omega}
      \label{eq:sensing_gain}
    \end{align}
\noindent Here $P_{\text{arm}}$ is the circulating arm power, $k$ the laser
wavenumber, $\Omega/2\pi$ the sideband
frequency of the GW readout, and $m$ each mirror mass. $L$ is the arm length of
$3995$ m and $\gamma$ the signal bandwidth of $2\pi {\cdot} 450$ Hz in LIGO. $G(\Omega)$ is
the optical field transmissivity between the arm cavities and readout detector,
making $2kG(\Omega)\sqrt{P_{\text{arm}}}$ the sensing function relating
$\delta x$ to the emitted optical field that modulates the homodyne readout power.

\begin{figure*}[bth]
\centering
\includegraphics[width=\linewidth]{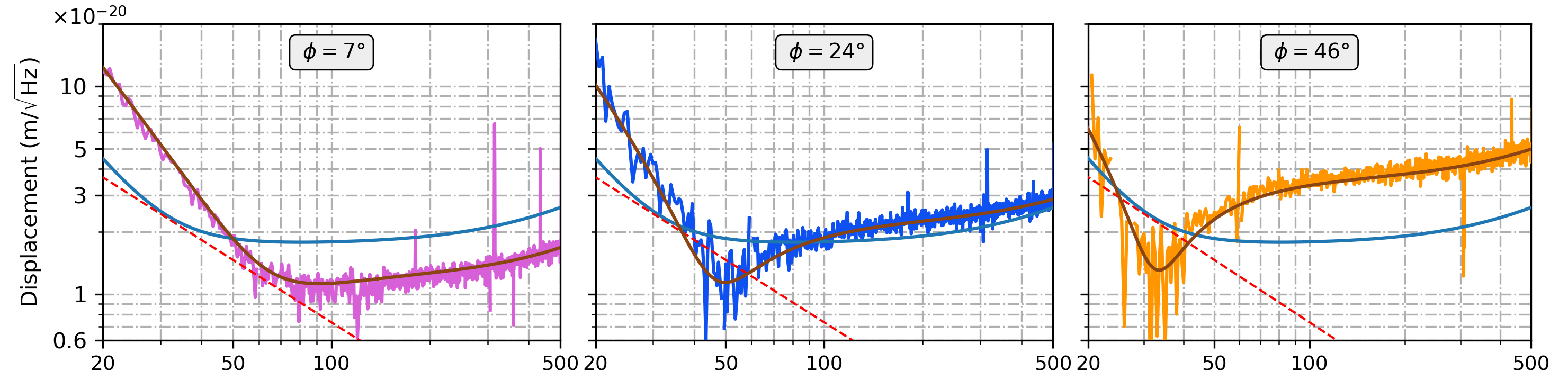}
\caption{Quantum noise spectra at additional squeezing angles of $7^\circ$ (magenta), $24^\circ$ (blue), $46^\circ$ (orange). Each data set is plotted with the same classical noise subtraction as Fig. \ref{fig:sub-SQLQRPN}, and with a corresponding quantum noise model curve (brown). The model without injected squeezing (blue) is plotted for comparison. The differences between the squeezed data sets and the reference model show that QRPN contributes to the motion of the Advanced LIGO mirrors. The QRPN contribution can be increased and decreased as the injected state is varied.  These data use less observing time than Fig. \ref{fig:sub-SQLQRPN} and have correspondingly larger statistical fluctuations.\label{fig:more_sqz_rot}}
\end{figure*}

\topic{ Squeezing factors:} The factors $S(\Omega, \phi)$ and ${(1 +
\mathcal{K}^2(\Omega))}$ capture the radiation pressure interaction whereby the
mirror oscillator motion correlates the injected optical amplitude quadrature to
the output phase quadrature, with $\mathcal{K}(\Omega)$ the pondermotive
interaction strength. The theory of pondermotive squeezing is detailed in Sec. IVA-B of \cite{KLMTV}. $S(\Omega, \phi)$ accounts for
injection of squeezed states. Without injected squeezing, $S {=} 1$,
in which case the arm power $P_{\text{arm}}$ may be chosen to minimize $\Delta
x(\Omega)$ by balancing shot noise and radiation pressure noise. The resulting
minimum $\Delta x(\Omega)$ is the free-mass SQL for a Michelson interferometer
with a Fabry-Perot cavity in each arm~\cite{KLMTV}:
    \begin{align}
       \Delta x^2(\Omega) &\ge \Delta x^2_{\text{SQL}}(\Omega)  \equiv \frac{8\hbar}{m\Omega^2}
                                          \label{eq:freemass_SQL}
    \end{align}
When injecting squeezed states
at squeeze angle $\phi$ with squeeze factor $r$, the squeezing measured at the readout,
$S(\Omega, \phi)$, becomes:
%
    \begin{align}
       S(\Omega, \phi) &= e^{-2r}\cos^2{\big(}\phi - \theta(\Omega){\big)}
          {+} e^{2r}\sin^2{\big(}\phi - \theta(\Omega){\big)}
          \label{eq:QL_sqz}
          \\
          \theta(\Omega) &= \arctan(\mathcal{K}(\Omega))
          \label{eq:theta}
    \end{align}
$\phi{=}0$ is defined as the squeezing angle that reduces the shot noise power spectral density, where $\theta {\rightarrow} 0$, by a factor of $e^{-2r}$.

\topic{ From $S(\Omega, \phi)$ to the dip:}
The expression $\phi {-} \theta(\Omega)$ characterizes the frequency-dependent interaction between pondermotive
and injected squeezing. Eqn.~\ref{eq:QL_sqz} indicates that at frequencies
where $\theta(\Omega) {=} \phi$, the two conspire to produce a minimum in the
quantum noise spectrum, appearing as a ``dip'' in the curves of
Fig.~\ref{fig:sub-SQLQRPN}. Whereas the $S = 1$ case led to the SQL in
Eqn~\ref{eq:freemass_SQL}, injecting squeezed states allows the SQL to be
surpassed at measurement frequencies for which $S(\Omega, \phi) < 1$.

\topic{ The measurement method:}
Fig~\ref{fig:sub-SQLQRPN} shows amplitude spectral densities of differential displacement. Exposing the sub-SQL dip requires reliably estimating and subtracting classical noise around \dipfrequency. The data are acquired as three sets of spectral measurements in each of
two operating modes -- with and without squeezing injection. By alternating operation between the two modes, we establish that the noise is consistent within statistical variations,
confirming that it is stationary over the duration of the experiment. To further address the concern that the classical noise between modes of operation may be changing, additional data at a range of squeezing angles are obtained, as shown in Fig.~\figsqzrotno.

\topic{ The reference measurement:} In Fig.~\ref{fig:sub-SQLQRPN}, the black trace
is the measured total noise at the readout with squeezing disengaged, including
both quantum and classical noise contributions. It is generated from a 90-minute
average split across three non-contiguous time periods where the squeezer cavity is
set off-resonance \cite{O3Squeezing}, allowing the unsqueezed vacuum state to enter the
interferometer. The blue trace is the modeled quantum noise contribution to the
total noise measurement of the black trace. Subtracting the blue trace from the
black trace gives the total classical noise contribution. We verify that
this classical noise component is stationary, and independent of squeezer status
(see discussion of Figure \figsqzrotno below and details in Methods). The model shows that quantum noise dominates the interferometer
sensitivity at high frequencies ($\Omega > \gamma \approx 2\pi {\cdot} 450$~Hz),
and accounts for \quantumnoiseatdip\ of the total measured noise power at \dipfrequency. Of the remaining non-quantum noise, \coatingandthermoopticnoise\ is estimated to be
coating and thermooptic noise, with the rest unidentified~\cite{instrO3paper}.

\topic{ The principle measurements:} The green trace of Fig.~\ref{fig:sub-SQLQRPN} shows the inferred quantum noise
spectrum with squeezing injected at $\phi {=} \sqzrotationforSQL$. 
This angle, determined from the model fit, places the dip in the frequency region where the ratio of the 
total noise in the reference spectrum and the SQL curve is minimized. The green trace is
calculated as the total measured displacement spectrum while the squeezer is
engaged, minus the classical noise contribution previously established from the
reference measurement. The purple trace shows the quantum noise model
corresponding to $\phi {=} \sqzrotationforSQL$ squeezing, featuring a dip in the
quantum noise that reaches down to \subsqlpercent\ or \subsqldB\ of the SQL at \dipfrequency.


\topic{ The extended measurements:} Squeezing measurements at three additional $\phi$'s are presented in Fig.\figsqzrotno. They show that QRPN contributes to the motion of the Advanced LIGO mirrors. 
 At each $\phi$, the quantum noise trace is calculated by
subtracting the same classical noise contribution (determined from the reference data)
from the measured displacement spectrum. We note that the modeled quantum noise
plotted here requires the full functional form of $S(\Omega,\phi,\psi)$
in Eqn.~\ref{eq:QL_SQZ_SRCL} in Methods, rather than the
simplified version of Eqn.~\ref{eq:QL_sqz}. These
additional measurements characterize contributions from an unwanted phase shift
due a slight detuning of the signal cavity, which manifests as a squeeze angle shift of
$\psi{=}\sqzrotationforSRCL$ accumulating across the frequency region where
$\Omega \sim \gamma$. A total of \numberofmeasurementsincontour\ squeezing measurements are
combined to plot $S(\Omega,\phi,\psi)$ in the Extended Data.

\topic{ Data and model uncertainties:}
Uncertainty in both data and model are
calculated here, with additional details in Methods. The statistical error in
the power spectrum measurement of the quantum noise, after subtraction, is
\statisticalerrorinqnpowerspec\ at \dipfrequency\ (for a
\binwidthforpowerspectrum\, bin width). We test for discrepancies between the
three reference datasets, and find that the relative uncertainty in the
classical noise stationarity is bounded by the same statistical error. Errors in
the optical sensing function $2kG(\Omega)\sqrt{P_{\text{arm}}}$, along with the
$\Delta x$ servo loop compensation, are determined from the online
interferometer calibration procedure \cite{instrO3paper}, and bounded to be
\calibrationuncertainty~\cite{Cal_O2}. Uncertainties in arm cavity power is
\armpoweruncertainty. Aside from the reference, the model curves of Figs. 2 and
3 require the squeeze factor $r$ and interferometer losses \cite{O3Squeezing},
which are determined from fits across all datasets, along with the
signal-recycling cavity detuning $\psi{=}\sqzrotationforSRCL$. Optical spring
effects are accounted in the calibration but, at this $\psi$, are insignificant
for the quantum noise model.

\topic{ GW and QM milestones:} 
The measurements presented here represent long-awaited milestones in
verifying the role of quantum mechanics in limiting the measurement of small
displacements generally, and in the sensitivity of GW detectors in particular.

\topic{ Backaction matters:}
First, we observe that QRPN contributes to the motion of the
kilogram-scale mirrors of LIGO. This observation is also made in the Advanced Virgo GW detector \cite{VirgoQRPN}. It is remarkable that quantum vacuum
fluctuations can influence the motion of these macroscopic, human-scale objects,
and that the effect is measured. This is quantum mechanics at its
experimentally most macroscopic scale.

\topic{ Backaction can be evaded:} Second, revealing quantum noise below the SQL in the Advanced LIGO detector is the
first realization of a quantum nondemolition technique in GW detectors \cite{SQL1,SQL2}, where quantum correlations prevent the measurement device from demolishing the same information one is trying to extract. Exploiting quantum correlations allows a fundamental quantum limit to be manipulated to improve measurement precision.

\topic{ Back to astrophysics:} Finally, we must not forget the foremost scientific objectives of the Advanced
LIGO detectors: they are designed for astrophysical observations of GWs from
violent cosmic events. During the third observing run, the squeezing angle is set to optimize the sensitivity to GWs from binary neutron star mergers~\cite{O3Squeezing}. This is not the
squeeze angle where shot noise is minimized, but where the combination
of shot noise and QRPN are minimized, implying that backaction evasion plays a
role in optimizing the sensitivity of the Advanced LIGO detector.
This is one of the
factors that has allowed Advanced LIGO to go from detecting roughly one
astrophysical event per month in observing runs 1 and 2, to about one
astrophysical trigger per week in the third observing run. 
In the future, with further mitigation of classical noise, sub-SQL performance of GW detectors promises ever greater astrophysical reach.




 
\bibliographystyle{naturemag}
\medskip
\bibliography{reference,refs_nergis}

\begin{thebibliography}{10}
\expandafter\ifx\csname url\endcsname\relax
  \def\url#1{\texttt{#1}}\fi
\expandafter\ifx\csname urlprefix\endcsname\relax\def\urlprefix{URL }\fi
\providecommand{\bibinfo}[2]{#2}
\providecommand{\eprint}[2][]{\url{#2}}

\bibitem{CavesPRD1981}
\bibinfo{author}{Caves, C.~M.}
\newblock \bibinfo{title}{Quantum-mechanical noise in an interferometer}.
\newblock \emph{\bibinfo{journal}{Phys. Rev. D}} \textbf{\bibinfo{volume}{23}},
  \bibinfo{pages}{1693--1708} (\bibinfo{year}{1981}).
\newblock \urlprefix\url{https://link.aps.org/doi/10.1103/PhysRevD.23.1693}.

\bibitem{SQL1}
\bibinfo{author}{Braginsky, V.~B.} \& \bibinfo{author}{Khalili, F.~Y.}
\newblock \bibinfo{title}{Quantum nondemolition measurements: the route from
  toys to tools}.
\newblock \emph{\bibinfo{journal}{Rev. Mod. Phys.}}
  \textbf{\bibinfo{volume}{68}}, \bibinfo{pages}{1--11} (\bibinfo{year}{1996}).
\newblock \urlprefix\url{https://link.aps.org/doi/10.1103/RevModPhys.68.1}.

\bibitem{SQL2}
\bibinfo{author}{Braginsky, V.~B.}, \bibinfo{author}{Khalili, F.~Y.} \&
  \bibinfo{author}{Thorne, K.~S.}
\newblock \emph{\bibinfo{title}{Quantum Measurement}}
  (\bibinfo{publisher}{Cambridge University Press}, \bibinfo{year}{1992}).

\bibitem{KLMTV}
\bibinfo{author}{Kimble, H.~J.}, \bibinfo{author}{Levin, Y.},
  \bibinfo{author}{Matsko, A.~B.}, \bibinfo{author}{Thorne, K.~S.} \&
  \bibinfo{author}{Vyatchanin, S.~P.}
\newblock \bibinfo{title}{Conversion of conventional gravitational-wave
  interferometers into quantum nondemolition interferometers by modifying their
  input and/or output optics}.
\newblock \emph{\bibinfo{journal}{Physical Review D}}
  \textbf{\bibinfo{volume}{65}}, \bibinfo{pages}{022002}
  (\bibinfo{year}{2001}).
\newblock \urlprefix\url{http://link.aps.org/doi/10.1103/PhysRevD.65.022002}.

\bibitem{Unruh1982}
\bibinfo{author}{Unruh, W.~G.}
\newblock \emph{\bibinfo{title}{Quantum Optics, Experimental Gravitation, and
  Measurement Theory}} (\bibinfo{publisher}{Plenum}, \bibinfo{year}{1982}).

\bibitem{BnC1}
\bibinfo{author}{Buonanno, A.} \& \bibinfo{author}{Chen, Y.}
\newblock \bibinfo{title}{Quantum noise in second generation, signal-recycled
  laser interferometric gravitational-wave detectors}.
\newblock \emph{\bibinfo{journal}{Phys. Rev. D}} \textbf{\bibinfo{volume}{64}}
  (\bibinfo{year}{2001}).

\bibitem{KweePRD2014}
\bibinfo{author}{Kwee, P.}, \bibinfo{author}{Miller, J.},
  \bibinfo{author}{Isogai, T.}, \bibinfo{author}{Barsotti, L.} \&
  \bibinfo{author}{Evans, M.}
\newblock \bibinfo{title}{Decoherence and degradation of squeezed states in
  quantum filter cavities}.
\newblock \emph{\bibinfo{journal}{Phys. Rev. D}} \textbf{\bibinfo{volume}{90}},
  \bibinfo{pages}{062006} (\bibinfo{year}{2014}).
\newblock \urlprefix\url{https://link.aps.org/doi/10.1103/PhysRevD.90.062006}.

\bibitem{Danilishin2008}
\bibinfo{author}{Danilishin, S.} \emph{et~al.}
\newblock \bibinfo{title}{Creation of a quantum oscillator by classical
  control}  (\bibinfo{year}{2008}).

\bibitem{PurduePhysRevD2002}
\bibinfo{author}{Purdue, P.} \& \bibinfo{author}{Chen, Y.}
\newblock \bibinfo{title}{Practical speed meter designs for quantum
  nondemolition gravitational-wave interferometers}.
\newblock \emph{\bibinfo{journal}{Phys. Rev. D}} \textbf{\bibinfo{volume}{66}},
  \bibinfo{pages}{122004} (\bibinfo{year}{2002}).
\newblock \urlprefix\url{https://link.aps.org/doi/10.1103/PhysRevD.66.122004}.

\bibitem{polzikBAE2017}
\bibinfo{author}{M{\o}ller, C.~B.} \emph{et~al.}
\newblock \bibinfo{title}{Quantum back-action-evading measurement of motion in
  a negative mass reference frame}.
\newblock \emph{\bibinfo{journal}{Nature}} \textbf{\bibinfo{volume}{547}},
  \bibinfo{pages}{191--195} (\bibinfo{year}{2017}).
\newblock
  \urlprefix\url{http://www.nature.com/nature/journal/v547/n7662/full/nature22980.html}.

\bibitem{qrpBoulder2013}
\bibinfo{author}{Purdy, T.~P.}, \bibinfo{author}{Peterson, R.~W.} \&
  \bibinfo{author}{Regal, C.~a.}
\newblock \bibinfo{title}{Observation of {Radiation} {Pressure} {Shot} {Noise}
  on a {Macroscopic} {Object}}.
\newblock \emph{\bibinfo{journal}{Science}} \textbf{\bibinfo{volume}{339}},
  \bibinfo{pages}{801--804} (\bibinfo{year}{2013}).
\newblock
  \urlprefix\url{http://www.sciencemag.org/cgi/doi/10.1126/science.1231282}.

\bibitem{schwabBAE2014}
\bibinfo{author}{Suh, J.} \emph{et~al.}
\newblock \bibinfo{title}{Mechanically detecting and avoiding the quantum
  fluctuations of a microwave field}.
\newblock \emph{\bibinfo{journal}{Science}} \textbf{\bibinfo{volume}{344}},
  \bibinfo{pages}{1262--1265} (\bibinfo{year}{2014}).
\newblock \urlprefix\url{http://www.sciencemag.org/content/344/6189/1262}.

\bibitem{wilsonNature2015}
\bibinfo{author}{Wilson, D.~J.} \emph{et~al.}
\newblock \bibinfo{title}{Measurement-based control of a mechanical oscillator
  at its thermal decoherence rate}.
\newblock \emph{\bibinfo{journal}{Nature}} \textbf{\bibinfo{volume}{524}},
  \bibinfo{pages}{325--329} (\bibinfo{year}{2015}).
\newblock
  \urlprefix\url{http://www.nature.com/nature/journal/v524/n7565/abs/nature14672.html}.

\bibitem{teufelPRL2016}
\bibinfo{author}{Teufel, J.}, \bibinfo{author}{Lecocq, F.} \&
  \bibinfo{author}{Simmonds, R.}
\newblock \bibinfo{title}{Overwhelming {Thermomechanical} {Motion} with
  {Microwave} {Radiation} {Pressure} {Shot} {Noise}}.
\newblock \emph{\bibinfo{journal}{Physical Review Letters}}
  \textbf{\bibinfo{volume}{116}}, \bibinfo{pages}{013602}
  (\bibinfo{year}{2016}).
\newblock
  \urlprefix\url{http://link.aps.org/doi/10.1103/PhysRevLett.116.013602}.

\bibitem{cripeNature2019}
\bibinfo{author}{Cripe, J.} \emph{et~al.}
\newblock \bibinfo{title}{Measurement of quantum back action in the audio band
  at room temperature}.
\newblock \emph{\bibinfo{journal}{Nature}} \textbf{\bibinfo{volume}{568}},
  \bibinfo{pages}{364--367} (\bibinfo{year}{2019}).
\newblock \urlprefix\url{https://www.nature.com/articles/s41586-019-1051-4}.

\bibitem{sudhirPRX2017}
\bibinfo{author}{Sudhir, V.} \emph{et~al.}
\newblock \bibinfo{title}{Quantum {Correlations} of {Light} from a
  {Room}-{Temperature} {Mechanical} {Oscillator}}.
\newblock \emph{\bibinfo{journal}{Physical Review X}}
  \textbf{\bibinfo{volume}{7}}, \bibinfo{pages}{031055} (\bibinfo{year}{2017}).
\newblock \urlprefix\url{https://link.aps.org/doi/10.1103/PhysRevX.7.031055}.

\bibitem{teufelNatureNano2009}
\bibinfo{author}{Teufel, J.~D.}, \bibinfo{author}{Donner, T.},
  \bibinfo{author}{Castellanos-Beltran, M.~A.}, \bibinfo{author}{Harlow, J.~W.}
  \& \bibinfo{author}{Lehnert, K.~W.}
\newblock \bibinfo{title}{Nanomechanical motion measured with an imprecision
  below that at the standard quantum limit}.
\newblock \emph{\bibinfo{journal}{Nature Nanotechnology}}
  \textbf{\bibinfo{volume}{4}}, \bibinfo{pages}{820--823}
  (\bibinfo{year}{2009}).

\bibitem{MasonNaturePhysics2019}
\bibinfo{author}{Mason, D.}, \bibinfo{author}{Chen, J.},
  \bibinfo{author}{Rossi, M.}, \bibinfo{author}{Tsaturyan, Y.} \&
  \bibinfo{author}{Schliesser, A.}
\newblock \bibinfo{title}{Continuous force and displacement measurement below
  the standard quantum limit}.
\newblock \emph{\bibinfo{journal}{Nature Physics}}
  \textbf{\bibinfo{volume}{15}}, \bibinfo{pages}{745--749}
  (\bibinfo{year}{2019}).
\newblock \urlprefix\url{https://doi.org/10.1038/s41567-019-0533-5}.

\bibitem{O3Squeezing}
\bibinfo{author}{Tse, M.}, \bibinfo{author}{Yu, H.},
  \bibinfo{author}{Kijbunchoo, N.} \emph{et~al.}
\newblock \bibinfo{title}{Quantum-enhanced advanced ligo detectors in the era
  of gravitational-wave astronomy}.
\newblock \emph{\bibinfo{journal}{Phys. Rev. Lett.}}
  \textbf{\bibinfo{volume}{123}}, \bibinfo{pages}{231107}
  (\bibinfo{year}{2019}).
\newblock
  \urlprefix\url{https://link.aps.org/doi/10.1103/PhysRevLett.123.231107}.

\bibitem{instrO3paper}
\bibinfo{author}{Buikema, A.} \emph{et~al.}
\newblock \bibinfo{title}{Sensitivity and performance of the advanced ligo
  detectors in the third observing run}.
\newblock \emph{\bibinfo{journal}{in preparation}}  (\bibinfo{year}{2019}).

\bibitem{O1instrPRL2016}
\bibinfo{author}{Abbott, B.~P.} \emph{et~al.}
\newblock \bibinfo{title}{Gw150914: The advanced ligo detectors in the era of
  first discoveries}.
\newblock \emph{\bibinfo{journal}{Phys. Rev. Lett.}}
  \textbf{\bibinfo{volume}{116}}, \bibinfo{pages}{131103}
  (\bibinfo{year}{2016}).
\newblock
  \urlprefix\url{https://link.aps.org/doi/10.1103/PhysRevLett.116.131103}.

\bibitem{Cal_O2}
\bibinfo{author}{Cahillane, C.} \emph{et~al.}
\newblock \bibinfo{title}{Calibration uncertainty for advanced ligo's first and
  second observing runs}.
\newblock \emph{\bibinfo{journal}{Phys. Rev. D}} \textbf{\bibinfo{volume}{96}},
  \bibinfo{pages}{102001} (\bibinfo{year}{2017}).
\newblock \urlprefix\url{https://link.aps.org/doi/10.1103/PhysRevD.96.102001}.

\bibitem{VirgoQRPN}
\bibinfo{author}{Acernese, F.} \emph{et~al.}
\newblock \bibinfo{title}{Quantum back-action on kg-scale mirrors - observation
  of radiation pressure noise in the advanced virgo detector}.
\newblock \emph{\bibinfo{journal}{in preparation}}  (\bibinfo{year}{2020}).

\bibitem{IzumiCrossCorr2017}
\bibinfo{author}{Kiwamu, I.}
\newblock \bibinfo{title}{Time domain implementation of dcpd cross
  correlation}.
\newblock \bibinfo{type}{Tech. Rep.} (\bibinfo{year}{2017}).
\newblock \urlprefix\url{https://dcc.ligo.org/LIGO-T1700131}.

\end{thebibliography}
\newpage
\section*{\large{METHODS}} 
\label{sec:Methods}

This section expands on four topics related to the measurement:
a) the augmented model for a non-ideal interferometer,
b) measurement uncertainty,
c) quantum noise model uncertainty,
d) non-stationary noise uncertainty, and
e) the additional plots in Extended Data.

\topic{Non-Ideal model }
The model curves present in Figures 2-5 are calculated from the full
coupled-cavity equations of \cite{BnC1}, which are exact and omit only effects
from high-order transverse optical modes. The model provided by equations
\ref{eq:QL_full}-\ref{eq:theta} represents an idealized interferometer with
all cavities on resonance and no optical losses. Here we extend the model to
consider the dominant experimental deviations from the ideal case, without the
complexity of the exact equations. This extension includes imperfect input and
output efficiency, as well as the additional frequency-dependent effect on the
squeezing angle from the small, unintended phase shift within the
signal-recycling cavity. For the parameters of this paper, the following model
is accurate to {\color{\blue} 5\%} of the exact model quantum power spectral
density between {\color{\blue}10Hz to 100Hz}.

The input and output efficiency of the interferometer are introduced using two
new parameters, \Ein\  and \Eout\  respectively. The input efficiency represents
the total fractional coupling of optical power between the squeezer cavity and
the interferometer, and the output efficiency is the total from the
interferometer to the readout homodyne detector. They must be considered
separately due to differences in their interaction with QRPN, leading to the
expressions:
%
    \begin{align}
       \Delta x^2(\Omega)  &= S^*{\cdot}\left(1 + \Eout\mathcal{K}^2(\Omega)\right)\frac{\hbar c }{\Eout 8k |G(\Omega)|^2 P_{\text{arm}} }
       \label{eq:deltaxlossy}
      \\
      (1-\Eeff) &= (1-\Ein) + \frac{1}{1 + \mathcal{K}^2(\Omega)} (1 - \Eout)
              \label{eq:loss_total}
      \\
      S^*(\Omega, \phi, \psi) &= \Eeff S(\Omega, \phi, \psi) + (1 - \Eeff)
              \label{eq:QL_SQZ_star}
            \\
      S(\Omega, \phi, \psi) &=
            e^{-2r}\cos^2{\big(}\phi - \theta^*{\big)}
          {+} e^{2r}\sin^2{\big(}\phi - \theta^*{\big)}
          \label{eq:QL_SQZ_SRCL}
          \\
          \theta^* &= \arctan(\mathcal{K}(\Omega)) + \frac{\Omega^2}{\gamma^2 + \Omega^2}\psi
          \label{eq:theta_star}
    \end{align}
%
External output loss does not change the dark-port to arm cavity optical field
transmissivity $G(\Omega)$, but it does modify the dark-port to readout
transmissivity, lowering the sensing function to be $2 k G(\Omega) \sqrt{\Eout
P_{\text{arm}}}$. This leads to the \Eout\ terms in Eqn. \ref{eq:deltaxlossy},
where shot-noise scales as $1/\Eout$, but the QRPN term does not. QRPN pertains
to real motion, and its reduced influence on the optical quantum noise is
compensated by the $\Delta x$ calibration.

A frequency-dependent effective efficiency, $\Eeff$, accounts for the output
loss $1 - \Eout$\  not being able to affect the real motion of the masses due to
radiation pressure, while the squeezed state is degraded by both input and
output losses. The form of Eqn. \ref{eq:loss_total} reflects the relation of
the input, output and effective losses rather than efficiencies, and it is
accurate for small losses.

The total squeezing angle shift due to the signal recycling cavity is encoded in
the parameter $\psi$. It appears alongside the pondermotive effect on the
squeezing angle in Eqn.~(\ref{eq:theta_star}), except it accumulates through the
cavity pole transition. This formulation is accurate for small detunings of the
interferometer signal cavity, and is related to the physical phase shift $\xi$
within the signal recycling cavity by $\psi{=}{\color{\blue} 10.7}\xi$, calculated
for the LIGO Livingston mirror parameters. Notably absent from this non-ideal
model but present in \cite{BnC1}, is the contribution of the optical-spring
effect due to $\xi{\ne}0$. We note that the above non-ideal model is accurate to
1\% in the zero detuning case ${\psi}=\xi{=}0$. While strong optical springs are
an alternative method of achieving sub-SQL quantum noise sensitivity, the
accuracy of the above augmented model indicates that the spring contribution is
mostly negligible for this measurement.


\topic{Uncertainties Overview}
Figure \ref{fig:sub-SQLQRPN} shows that quantum noise accounts for only
\quantumnoiseatdip\ of the total interferometer noise power at \dipfrequency.
For this reason, classical noises must be subtracted in order to reveal the
quantum limited displacement sensitivity. The interferometer is a complex
instrument with such sensitivity that the following considerations must be
addressed to validate the subtraction. First, the fiducial quantum noise model
of the reference dataset and the parameters it relies on must be established and
the data must be calibrated. Second, the classical noise established for the
reference operating mode must be representative of the classical noise during
squeezing operation. In particular, the classical noise during the reference
period must not be higher than during squeezing, which would bias our inference
to underestimate the quantum noise contribution during squeezing. The reference
and squeezing datasets are taken in multiple, alternating segments and we test
for variations arising from non-stationary (time-varying) noise. Furthermore,
the non-stationary noise power contributions are mitigated by using a
statistically ``robust'' median based computation to calculate the sampled power
spectra.

\topic{Methods Outline}
The following paragraphs proceed by detailing the measurement sequence used to
characterize the stationarity, then describe how uncertainty propagates
through the data analysis for the post-subtraction quantum noise curves. We then
show how the calibration and interferometer data outputs are combined with
external measurements to establish our quantum noise model. The statistical
uncertainty is then outlined, followed by the methodology for characterizing the
noise stationarity between squeezing and reference datasets. Finally, the
spectral density estimator is described.

\topic{Measurement Sequence}
The data shown in Fig. \ref{fig:sub-SQLQRPN} were taken over a 5 hour
period on the advanced LIGO detector. To avoid variations of classical noise and
calibration, the interferometer power is held constant across all measurements.
To minimize statistical error, the majority of the measurement time is spent in
the two modes plotted: three 30-minute ``reference'' segments with the squeezer
disabled, alternating with three 30-minute segments with squeezing at $\phi{=}
\sqzrotationforSQL$. Each reference segment is following by a squeezing segment,
alternating three times to establish that the classical noise contribution is
constant across the total duration. The remaining time is split across nine
additional segments at varying input squeezing angles, and the final segment is
a fourth reference without squeezing.

\topic{ Uncertainty model}
Here we describe how the uncertainty propagates through the subtraction in our
measured quantum noise curves. The symbols for the frequency dependent reference
and squeezing data are $D_r$, $D_s$, and $M_r$, $M_s$ for the models. The
post-subtraction inferred quantum noise is given as $Q$ in the expression
%
    \begin{align}
      Q(\Omega) &= D_s(\Omega) - \big(D_r(\Omega) - M_r(\Omega)\big)
    \end{align}
%
The relative error of the post-subtraction squeezed quantum noise is given by
$\delta Q$, composed of the quadrature sum of relative errors due to the optical
sensitivity calibration, $\delta G$; the servo loop calibration, $\delta C$; the
modeling uncertainty, $\delta M_r$; statistical fluctuations $\delta D_r$,
$\delta D_s$; and relative stationarity uncertainty terms, $\delta N_t$, and
$\delta N_m$. All of these uncertainties are frequency-dependent, but the
argument $\Omega$ is suppressed for space. The definitions of these components
are clarified in the text following, but contribute to the expression:
%
    \begin{align}
      \delta Q^2 &=
                   \delta G^2 +
                   \frac{1}{Q^2}\Big(
                   M_r^2{\cdot}\delta M_r^2
                   + (D_r - D_s)^2{\cdot}\delta C^2
                   \nonumber\\
                 &\hspace{3ex}
                   +  D_r^2{\cdot}\delta D_r^2
                   +  D_s^2{\cdot}\delta D_s^2
                   \nonumber\\
                 &\hspace{3ex}
                   + (D_r - M_r)^2{\cdot}(\delta N_t^2 + \delta N_m^2)
                   \Big)
                   \label{eq:uncert_model}
    \end{align}
%
The lines of the above relation represent terms with different magnitudes of scaling terms. Given that \mbox{$Q \approx M_s \sim D_r - D_s$}, the top line for the calibration and model error has terms with order-1 coefficients, indicating that the relative errors quoted in the main text remain small for the comparison to the dip model. The lower two lines of eq. \ref{eq:uncert_model} show that the relative statistical fluctuations and stationarity uncertainties are magnified by the ratio, $V$, of the total classical PSD to the squeezed quantum PSD, approximately a factor of $V {=} \sqlrelativetotal$, at \dipfrequency.

\topic{ Uncertainty in quantum noise:}
The first line of Eqn. \ref{eq:uncert_model} includes the calibration and
unsqueezed reference quantum noise model uncertainty terms, $\delta G$, $\delta
C$, $\delta M_r$. The LIGO online calibration system determines the optical
sensing function $2 k G(\Omega) \sqrt{\Eout P_{\text{arm}}}$ which affects both
the model and calibration uncertainties. To prevent double-counting in the
incoherent sum, this optical gain has been isolated to the factor $\delta G$ and
should not be considered in $\delta C$ or $\delta M_r$. The sensing function is
monitored continuously by injecting displacement signals at several frequencies.
Some of these appear as narrow lines in the measured spectra of Figure
\ref{fig:sub-SQLQRPN}. From these continuous injections, the bandwidth $\gamma$
and the product $\Eout P_{\text{arm}}$, are determined. In addition, parameters
related to the optical spring are measured \cite{Cal_O2}, but primarily affect the
sensing function at frequencies {\color{\blue} ${<}10$Hz} for the measured
detuning. Additional lines monitor the $\Delta x$ servo loop actuators to apply
the frequency-dependent correction for the servo closed loop response, which is
contained in $\delta C$. The quoted calibration uncertainty of
\calibrationuncertainty\ is the incoherent sum $\sqrt{\delta G^2 + \delta
C^2}$.

\topic{ $\delta M_r$ Components }
Having factored $\delta G$ out of $\delta M_r$, any error in subtracting the
classical noise estimate between reference data and model can only arise from
estimating the shot noise and QRPN components represented by the term $g(1 + \Eout \mathcal{K}^2(\Omega))$. Here, $g$ is a scale factor relating homodyne power to
optical field. It is unknown because the calibration system exports its sensing
function in an end-to-end fashion with the photodectors in arbitrary voltage
digitization units; however, the $g$ may be well estimated using a
cross-correlation method detailed below. The remaining $g\mathcal{K}^2(\Omega)$
contribution may be estimated from the factors
$|G(\Omega)|^2\sqrt{P_{\text{arm}}}$. Independent measurements establish the
quoted arm power $P_{\text{arm}}{=}\circulatingarmpower$ kW, and this, combined
with the optical sensing gain calibration, allows us to determine the output
efficiency, $\Eout$. The squeezing level at high frequencies is determined by
$r$ and $\Eout{\cdot}\Ein$ (see Eqns.~\ref{eq:loss_total}-\ref{eq:QL_SQZ_star}),
and using the extended datasets with $\phi=0^\circ$, the input efficiency $\Ein$
may be determined from the observed readout squeezing level.

\begin{figure*}[htb]
\centering
\includegraphics[width=0.95\textwidth]{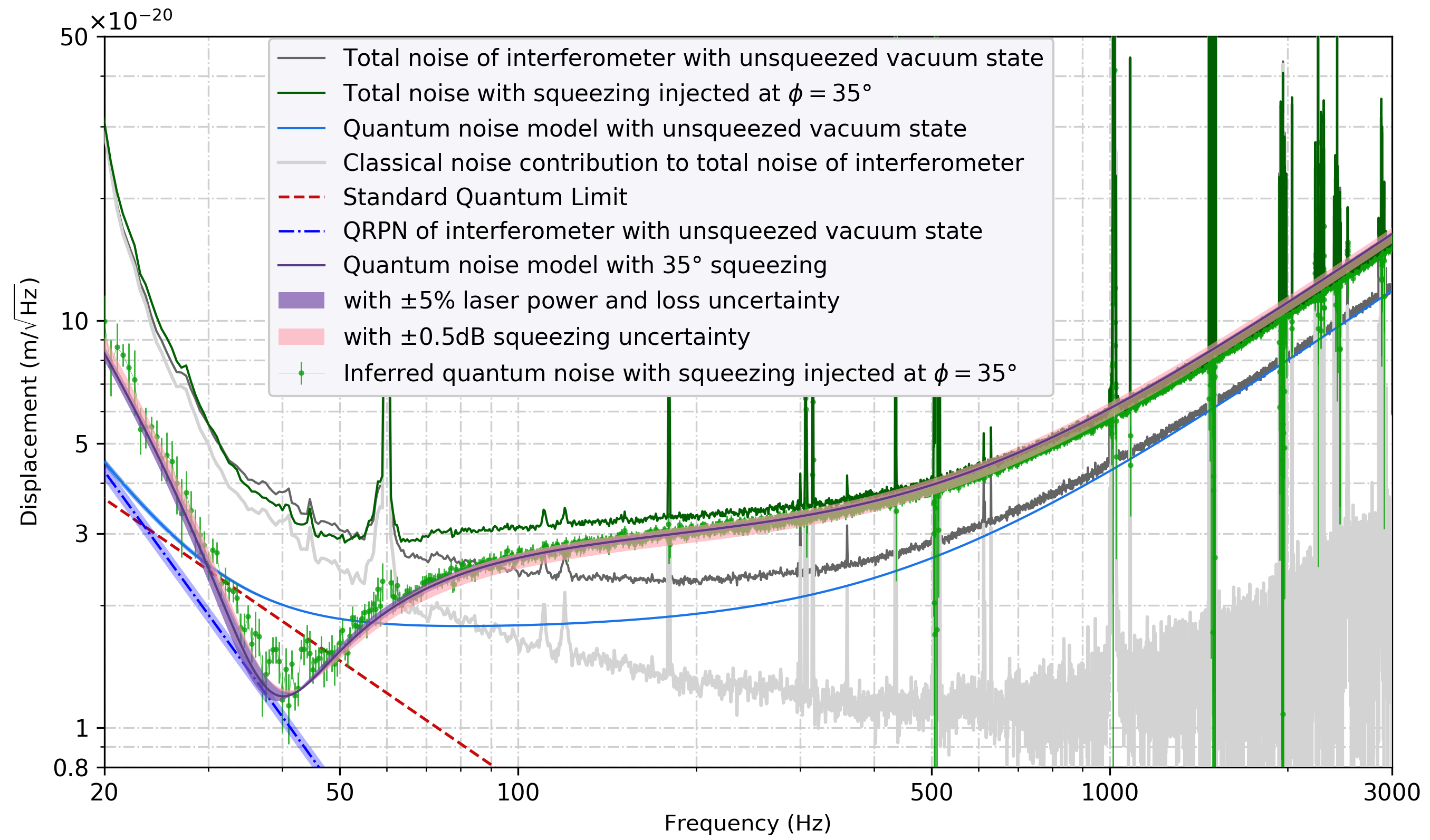}
\caption{The differential displacement ($\Delta x$) noise spectra density of the interferometer with uncertainties. The black and dark green trace show the measured total noise level of the interferometer with unsqueezed vacuum state (the reference) and injected squeezing at 35$^\circ$ respectively. The grey curve shows the classical noise contribution to the total noise of the interferometer, which is independent of the squeezer state. The solid blue quantum noise model curve includes the 5\% uncertainty in the arm power, but compensated by the output optical loss to model the calibrated sensing function. The green inferred quantum noise curve includes the statistical uncertainty of both the classical noise from the reference measurement, as well as in the squeezed measurement. The purple quantum noise model with 35$^\circ$ squeezing is shown with 5\% arm power uncertainty (purple shaded) and 0.5dB uncertainty of squeezing generated by the squeezer (pink shaded). The free-mass SQL is shown by the dashed red line, and the pure QRPN contribution of the interferometer with unsqueezed vacuum state is shown by the dashed blue line with its uncertainties from arm power.}
\label{fig:SQL_errorbar}
\end{figure*}

\topic{Modeling Uncertainty Plotted} The parameters describing the status of the
interferometer and squeezer during the experiment are listed in Table 1 of
Extended Data with uncertainties. They are also the values used in the modeling
of quantum noise calculation. Immediately before the 5 hour dataset, the
nonlinear parameter of the squeezer was measured to calculate $r$. The
squeezing angle is determined ultimately through a model fit, but it agrees with
our knowledge of the nonlinear conversion from the coherent control field
demodulation angle to the observed squeezing angle and the settings during the
shot-noise squeezing ($\phi{=}0^\circ$) and antisqueezing ($\phi{=}90^\circ$)
datasets. The frequency-dependent contributions of the squeezing and arm power
modeling uncertainties are shown in Fig.~\ref{fig:SQL_errorbar}, and they do not
strongly influence the model at the sub-SQL dip.

\topic{ Readout Cross Correlation}
The following cross-correlation method \cite{IzumiCrossCorr2017} is used to
determine the factor $g$, that relates the arbitrary experimental photodetector
units back to the physical optical field units. Two photodetectors are located
at the readout port of the LIGO interferometer (see figure
\ref{fig:instrument_layout}). When squeezing is not injected, shot-noise and
readout electronics noise (i.e. dark noise) are uncorrelated between the two
photodetectors, while QRPN and all of the classical noises are correlated. If
the cross correlation and dark noise is subtracted from total noise power for
the reference dataset, then only the shot noise remains, calibrated to
displacement. This precisely determines the optical sensing gain in physical
units, up to the uncertainty $\delta G$. The dark noise is only $1\%$ of the
shot noise power and so contributes negligibly to the uncertainty in this
subtraction.

\topic{ Statistical uncertainty}
The statistical uncertainty arises in that the fluctuations intrinsic to noise
also limit our ability to estimate it. With total measurement time $T_i$ for a
given dataset $i$, and bin width of \mbox{$\Delta F =
\mathrm{\binwidthforpowerspectrum}$}\ in the spectral density calculation, the
relative statistical uncertainty of the inferred quantum noise power is $\delta
D_i = (E T_i \Delta F)^{-1/2}$, with $E$ the statistical efficiency accounting
for the spectral estimation method. For the median method detailed below, we
determine through numerical experiments on white noise that $E=1.0$ for single-bin
error bars. The bin-bin covariance due to the apodization window causes
$E=60\%$ when averaging multiple adjacent datapoints. The total statistical
uncertainty of \statisticalerrorinqnpowerspec\ includes both datasets $\delta
D_r$ and $\delta D_s$ and their scaling by $V$ in Eqn.~\ref{eq:uncert_model}.

\topic{ Stationarity Uncertainty Definition}
Here we describe and characterize the terms $\delta N_t$, $\delta N_m$ in the
uncertainty budget of Eqn.~\ref{eq:uncert_model}. We label these terms together
the stationarity uncertainty, and they are intended to quantify potential
variations between the classical noise power as estimated from the unsqueezed
reference dataset and the classical noise power actually present in the
squeezing measurements. Under the presupposition that the models, $M_r$, and
$M_s$ are perfect, and the statistical noise is small, these uncertainties are
defined as the relative difference \mbox{$D_s {-} M_s \equiv (D_r {-} M_r){\cdot}(1
{+} \delta N_t {+} \delta N_m)$}. The two are distinguished as the changes to
classical noise arising from variations in time, $\delta N_t$, and from
switching the physical operating mode between the reference and squeezing, $\delta
N_m$.

\topic{ Stationarity Uncertainty Mitigation}
The time variation contribution to non stationarity, $\delta N_t$, is mitigated
both through the spectral density estimation method and the use of three
alternating segments for the reference and squeezed data. The aim of the
alternating segments is for the operating mode to switch on a timescale faster
than the environmental variation. The environmental timescale is not known or
even well-defined, so instead the discontiguous segments of reference time are
compared, setting a limit to the non-stationarity of the squeezing segment
between them. This is done likewise for the squeezing segments surrounding a
reference segment. We define a metric for the relative non-stationarity between
two such discontiguous segments to be
%
    \begin{align}
    \mathcal{N}_{ij} &= 2 \frac{D_i-D_j}{D_i+D_j}
    \end{align}
%
Each pair of datasets makes an estimate of the noise contribution varying at and
below the separation timescale of the datasets, here 1 hour. Each estimate
$\mathcal{N}_{ij}$ is limited by the statistical error of the constituent
datasets, and they are shown in Fig.~\ref{fig:subtraction_methods}. Since each
pair only constitutes a fraction of the full data, the multiple estimates are
combined to reduce the statistical uncertainty.
    \begin{align}
      \mathcal{N}_\Sigma^2
      &
        = \frac{1}{6}\big(
        \mathcal{N}_{R12}^2 + \mathcal{N}_{R23}^2 + \mathcal{N}_{R31}^2
        \nonumber \\&\hspace{3em}
      + \mathcal{N}_{S12}^2 + \mathcal{N}_{S23}^2 + \mathcal{N}_{S31}^2
      \big)
    \end{align}
%
Finally, these metrics must be related to the stationarity term $\delta N_t$.
The averaged nonstationary power $\mathcal{N}_\Sigma^2$ represents an estimate
of the time-varying contibution between adjacent reference and squeezing
segments, of which there are three. For many such segments, assuming random
fluctuations to the environmental noise level at the alternation time scale, the
contributions add in quadrature to give $\delta N_t^2 \lesssim
\mathcal{N}_{\Sigma}^2 / 3$. We then propagate the statistical noise limits for
segments one third the length of the total reference time $T$. This arrives at
the statistical limit to our stationarity uncertainty of $\delta
N_t{\approx}\sqrt{2}(E T \Delta F)^{-1/2}$. Because the total squeezing data
time is also $T$, our limit to the time variation contribution to
non-stationarity evaluates to be the same as the total statistical uncertainty
from both the squeezing and unsqueezed datasets, $\delta N_t^2 \approx \delta
D_r^2 + \delta D_s^2$. In addition to the individual
pairs, Fig.~\ref{fig:subtraction_methods} also shows the combined estimate
$\mathcal{N}_\Sigma^2$.

\begin{figure*}[bth]
\centering
\includegraphics[width=0.9\textwidth]{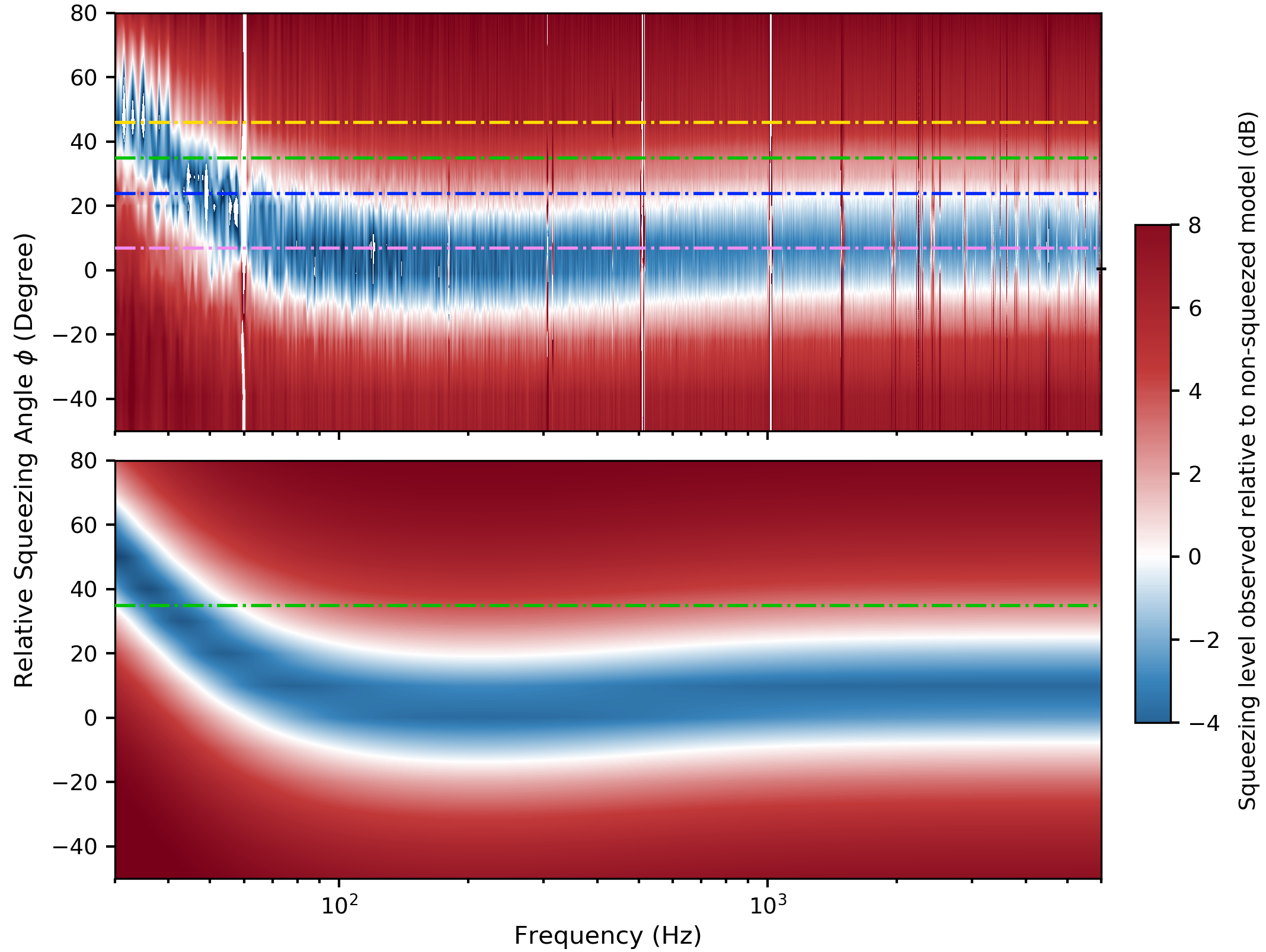}
\caption{
  {\bf Squeezing level and quantum noise of the interferometer over a
    full range of squeezing angles} Contour plot of squeezing level
  $S*(\phi, \theta, \psi)$ detected in the interferometer as a function of frequency and squeezing angle $\phi$
(upper), and its theoretical model (lower). Right: Quantum noise spectra at
selected squeezing angles. The contour plots show the frequency-dependent
squeezing level over a full range of squeezing angles from -50$^\circ$ to
80$^\circ$. Dashed lines crossing the plot represents $\phi =$ 35$^\circ$
(green), 7$^\circ$ (pink), 24$^\circ$ (blue) and 46$^\circ$ (orange),
corresponding to the quantum noise spectra of Figs.~\ref{fig:sub-SQLQRPN} and 3.
}
\label{fig:full_range_sqz_ang_rot}
\end{figure*}

\begin{figure*}[htb]
\centering
\includegraphics[width=0.9\textwidth]{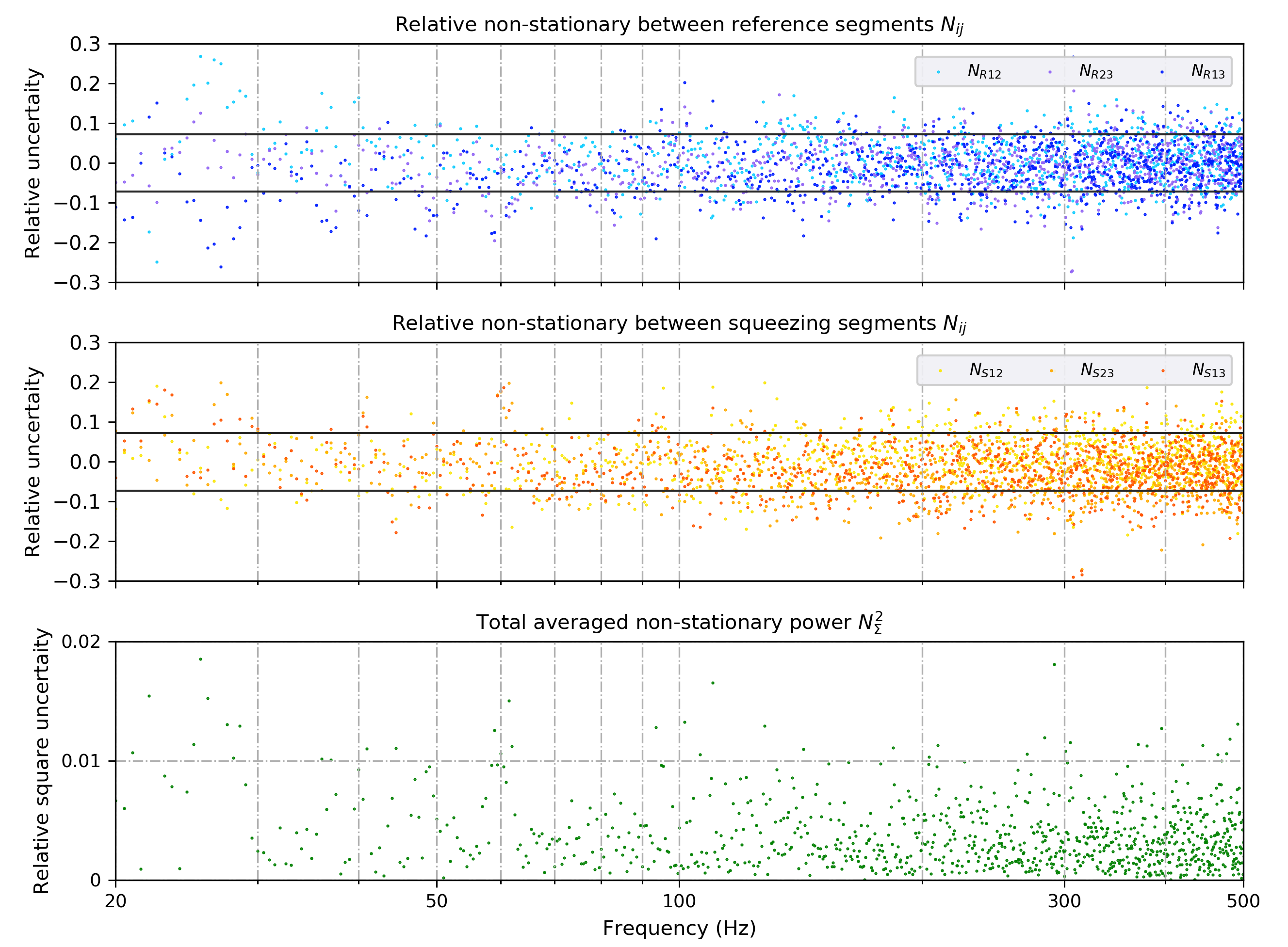}
\caption{Individual and combined estimates of non-stationary noise
between measurement segments | The upper two plots show the relative
time variation of noise between each pair of reference and squeezing measurement
segments, respectively. The black lines shown are 2$\sigma$ standing for 95\%
confidential level. The bottom plot shows the combined non-stationary power
defined by Eqn. 14.}
\label{fig:subtraction_methods}
\end{figure*}

The operating-mode varying component $\delta N_m$ of non-stationary noise is
constrained by the following arguments. The first is that it is quantitatively
constrained by the data at additional squeezing angles depicted in
Fig.~\figsqzrotno ~of the main text and
Fig.~\ref{fig:full_range_sqz_ang_rot} of the Extended Data. There, the same
classical noise estimate is subtracted and the model curves maintain their
agreement with the inferred quantum noise at alternate squeezing angles. Those
datasets however have limited statistical bounds due to their short duration.
The term $\delta N_m$ may be considered small for the following physical
reasons. The primary reason is that during the without-squeezing time, the optical
path is not changed, only the squeezer OPO is operated off of resonance to stop
its nonlinear parametric interaction. This means that environmental scatter
noise - the very low-power light leaking from the interferometer to the squeezer
system - does not impinge on different scattering surfaces between the two
modes. In the event that such scatter does matter, the fourth reference taken at
the end of the entire measurement period uses an in-vacuum beam diverter to
block the path to the squeezer. Testing that fourth reference against the other
three through the $\mathcal{N}_{ij}$ method shows no significant changes to the
classical noise.

In the event that the classical noise does change from the switch to squeezing,
we argue that the addition of the nonlinear parametric interaction from the
squeezer on this scattered light is more likely to increase the noise only
during the squeezing segments. This implies that the measurement should not be
biased low and will not over-estimate how much we have surpassed the SQL.
Indeed, the few data points in Figs.~\ref{fig:sub-SQLQRPN} and
\ref{fig:SQL_errorbar} that exceed the model beyond the statistical fluctuations
may be due to such a squeezer-specific noise source. We attribute the minimal
classical noise contribution to the use of a traveling wave OPO cavity,
in-vacuum suspended layout and coherent control implementation
\cite{O3Squeezing}.

\topic{ Spectral density estimation}
Finally, we describe the median method used for our spectral density estimation.
We claim through the above arguments that the classical noise is established to
be stationary in these datasets, however it is known from astrophysical analysis
that these complex detectors have intermittant time-resolved glitches and
artifacts of varying strengths. Intervals of excess noise are nontrivial to
identify due to the inherently random nature of noise, and time-resolved noise
power vetoes can introduce selection bias. We use the Welch - Bartlett overlap
method to estimate the power spectral density with no selection vetoes. Instead,
rather than averaging the individual spectra independently at each frequency,
the sample median at each frequency is taken. This generates a bin-by-bin median
strain spectral power density.

Initially, the entire period for a given spectral density estimate is split into
N 2-second segments, where each segment overlaps the segment before it by 50\%,
implementing the Welch method. For each segment, the time-series is linear
detrended and a Hann window is applied, then converted to a displacement
spectrum by a Fourier Transform. The collection of segments gives N estimates of
the power density in each frequency bin, each nominally following a chi-square
distribution on two variables (the real and imaginary parts of the Fourier
transform), but the distribution has an extended tail due to glitches and
transients of the detector. The median is picked for each frequency bin, and
then a computed scale factor is applied to convert the distribution median to
the mean noise power. This technique is unbiased for stationary noise, and
greatly improves the robustness to glitches and nonstationary contributions,
without selection bias from time-domain band-limited noise vetos. The downside
is that the statistical efficiency is approximately $\sqrt{2}$ worse than the
typical Welch method for a given spectrum averaging time.

\topic{ Extended Data Plots}
Fig. \ref{fig:SQL_errorbar} of Extended Data shows a variation of Figure 2
spanning a wider frequency range. The figure includes the
frequency-dependent uncertainties of Eqn. \ref{eq:uncert_model} in its model
curves and subtracted quantum noise plots.

Fig.~\ref{fig:full_range_sqz_ang_rot} shows a measurement of (upper) and model
of (lower) the squeezing term $S^*(\Omega,\phi,\psi)$ of the augmented model.
The quantum noise spectrum at ten additional $\phi$'s is determined by
subtracting the classical noise contribution (previously established through the
reference measurement) from the measured displacement spectrum at each
$\phi$. Each inferred quantum noise spectrum is then divided by the modeled
quantum noise spectrum without injected squeezing (blue trace in
Fig.\ref{fig:sub-SQLQRPN}) to obtain the observed squeezing term
$S^*(\Omega,\phi, \psi)$. The dashed lines indicate cross-sections in other
figures. Green is $\phi{=}\sqzrotationforSQL$ in Fig.~\ref{fig:sub-SQLQRPN}, and
yellow blue and purple correspond to the angles of Fig.~\figsqzrotno.


{\bf Acknowledgements:} LIGO was constructed by the California Institute of Technology and Massachusetts Institute of Technology with funding from the National Science Foundation, and operates under Cooperative Agreement No. PHY-0757058. Advanced LIGO was built under Grant No. PHY-0823459. The authors also gratefully acknowledge the support of the Australian Research Council under the ARC Centre of Excellence for Gravitational Wave Discovery, grant No. CE170100004 and Linkage Infrastructure, Equipment and Facilities grant No. LE170100217 and Discovery Early Career Award No. DE190100437; the National Science Foundation Graduate Research Fellowship under Grant No. 1122374; the Science and Technology Facilities Council of the United Kingdom, and the LIGO Scientific Collaboration Fellows program. 
\newpage


\begin{table}[t]
\centering
\resizebox{\linewidth}{!}{
\begin{tabular}{l lr }
 & &\\
\cline{1-3}
\textbf{Interferometer Parameter}   & Value\\
\cline{1-3}
Laser power in the arm cavity, ($P_{\text{arm}}$) & \circulatingarmpower\ kW\\
Optical loss before IFO, (1-$\Ein$)     & {17.2\%}\\
Optical loss after IFO, (1-$\Eout$)     & {17.4\% }\\
SRM phase detuning, ($\xi$)          & {$15$ mrad}\\
 \\
\cline{1-3}
\textbf{Squeezer Parameter} & Value\\
\cline{1-3}
Measured OPO nonlinear gain          & {\nlgforSQL\ }\\
Squeezing ideally generated by OPO ($e^{-2r}$) & {$9.8{\pm}0.15$ dB }\\
Squeezer phase noise ($\delta \phi$)  & {0-50 mrad}\\
Squeezing quadrature rotation angle ($\phi$)  & \sqzrotationforSQL\\
Max phase squeezing in IFO           & \measuredSQZforSQL\ \\
Max phase anti-squeezing in IFO       & \measuredASQZforSQL\ \\
\end{tabular}}

\caption{L1 Interferometer and Squeezer Parameters used for Modeling}
\label{tab:ifo_squeezer_parameters}
\end{table}


%
\section*{}


%

\newpage

\end{document}